# Orientation of the electric field gradient and ellipticity of the magnetic cycloid in multiferroic BiFeO$_3$

A. Pierzga[1], A. Błachowski[1], K. Komędera[1], K. Ruebenbauer[1*], A. Kalvane[2], and R. Bujakiewicz-Korońska[3]

[1]*Mössbauer Spectroscopy Division, Pedagogical University*
*PL-30-084 Kraków, ul. Podchorążych 2, Poland*

[2]*Institute of Solid State Physics, University of Latvia*
*Kengeraga 8, LV-1063, Riga, Latvia*

[3]*Material Physics Division, Pedagogical University*
*PL-30-084 Kraków, ul. Podchorążych 2, Poland*

[*]*Corresponding author:* sfrueben@cyf-kr.edu.pl

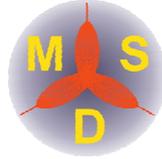

**Abstract**

The paper deals with the hyperfine interactions observed on the $^{57}$Fe nucleus in multiferroic BiFeO$_3$ by means of the 14.41-keV resonant transition in $^{57}$Fe, and for transmission geometry applied to the random powder sample. Spectra were obtained at 80 K, 190 K and at room temperature. It was found that iron occurs in the high spin trivalent state. Hyperfine magnetic field follows distribution due to the elliptic-like distortion of the magnetic cycloid. The long axis of the ellipse is oriented along $\langle 111 \rangle$ direction of the rhombohedral unit cell. The hyperfine magnetic field in this direction is about 1.013 of the field in the perpendicular direction at room temperature. This ratio diminishes to 1.010 at 80 K. Axially symmetric electric field gradient (EFG) on the iron atoms has the principal axis oriented in the same direction and the main component of the EFG is positive. Our results are consistent with the finding that iron magnetic moments are confined to the $[1\bar{2}1]$ crystal plane.

## 1. Introduction

The compound BiFeO$_3$ crystallizes in the cubic perovskite structure at high temperature, and behaves as an insulator. Ionic states of particular components are relatively well defined with bismuth occurring in the trivalent state with the lone pair of the predominantly 6s$^2$ character. Iron occurs in the octahedrally coordinated high spin trivalent state. However, some admixture of the covalent bonds between O$^{2-}$ anion and respective cations is observed. A rhombohedral distortion occurs at about 1100 K leading to the loss of the inversion center and development of the ferroelectricity [1]. The electric polarization vector is oriented along the $\langle 111 \rangle$ direction of the rhombohedral chemical unit cell, the latter cell having angle between main axes $\alpha = 89.375°$ and lattice constant $a = 0.39581$ nm at room temperature [2]. The lack of the inversion center is likely to be a combined effect of the oxygen octahedron distortion (rotation), iron atom displacement (along electric polarization vector) and presence of the stereo-active lone pair in the vicinity of the trivalent bismuth ions. Stereo-activity of the bismuth lone pair is induced by the lattice distortion. Oxygen octahedron rotation and iron atom displacement lead to differentiation of the iron - oxygen distances in the nearest neighbor shell of the oxygens surrounding iron. Oxygen octahedra rotate in the opposite way in adjacent chemical cells joint by the line along the $\langle 111 \rangle$ direction. Hence, two inequivalent iron sites are generated with the local symmetry being lower than cubic [2, 3], however they seem equivalent one each other while looking from the iron nucleus at the close neighborhood. The latter distortions lead to the creation of the electric field gradient (EFG) on the iron atomic nuclei. It is interesting to note that the compound starts to decompose even below the ferroelectric phase transition at least under ambient pressure at about ~870 K. Magnetic moments of iron order antiferromagnetically at about 653 K [4]. A magnetic structure due to the order of the trivalent high spin magnetic moments of iron is antiferromagnetic having the G-type character, i.e., each iron magnetic moment is surrounded by the nearest neighbor iron magnetic moments pointing in opposite direction. Magnetic moments make an incommensurate cycloid propagating in the $\langle 10\bar{1} \rangle$ direction. All magnetic moments are confined to the plane defined by the electric polarization vector $\langle 111 \rangle$ and cycloid propagation vector $\langle 10\bar{1} \rangle$, i.e. they are confined to the $[1\bar{2}1]$ plane of the rhombohedral chemical unit cell. A spiral magnetic structure is generated by the Dzyaloshinskii-Moriya interaction, and it has exceptionally large period, the latter being about 62 nm [5]. Hence, one can conclude that the orbital contribution to the iron magnetic moment is incompletely quenched. The Mössbauer spectra exhibit apparent two sub-spectra differing by the hyperfine field and electric field gradient [6]. Alternatively some "exotic" hyperfine field distribution could be fitted to the data [6-8]. Hence, it is interesting to have a closer look on the hyperfine interactions in this complex system with antiferromagnetically coupled (shifted in phase by $\pi$) long spin cycloids. The aim of this contribution is to attempt to resolve the problem of the Mössbauer spectrum shape by application of some physically feasible model.

## 2. Experimental

Polycrystalline sample in the powder form was prepared as described in [9]. Mössbauer spectra were collected for powder sample in a transmission mode using 14.41-keV line of $^{57}$Fe. The MsAa-3 spectrometer was used with the Kr-filled proportional counter and commercial $^{57}$Co(Rh) source kept at room temperature. The Janis Research Co. SVT-400TM cryostat was used to maintain the absorber temperature with the accuracy better than 0.01 K. Velocity scale was calibrated by the Michelson-Morley interferometer equipped with the He-Ne laser. Spectra were calibrated and processed by means of the proper applications belonging to the Mosgraf-2009 suite [10]. Spectra were fitted within standard transmission integral approximation by using GmFeAs application of the Mosgraf-2009 suite. All spectral shifts are reported versus room temperature $\alpha-Fe$.

## 3. Discussion of results

Mössbauer spectra are composed of four distinct components all generated by the high spin trivalent iron. Spectra are shown in Figure 1, while the essential parameters are gathered in Table 1. The symbol RT stands for the room temperature. The first (major) component belongs to the $BiFeO_3$ phase and makes about 89 % contribution to the resonant cross-section.

There are at least three impurity sites. Two of them order magnetically below room temperature, while the site denoted as the second does not order till 80 K. The quadrupole interaction on these sites is described in the same manner as for the major site. However, the first order approximation is used in place of the full Hamiltonian used for the major site. It is assumed that the EFG is axially symmetric for impurity sites. The sign of the quadrupole coupling constant remains undetermined for the non-magnetic sites.

The best fit to the major site is obtained introducing evenly spaced magnetic moments (hyperfine fields) confined to the plane and described by the expression $B(\phi) = B_0 \exp[P_{20}\cos^2\phi] \approx B_0[1 + P_{20}\cos^2\phi]$. The symbol $B_0 > 0$ denotes scaling field, while the symbol $P_{20}$ denotes departure from circularity of the cycloid, i.e. the cycloid ellipticity. One can assume that the angle $0 \leq \phi < 2\pi$ is confined to the $[1\bar{2}1]$ plane containing $\langle 111 \rangle$ direction, i.e., the electric polarization vector as well. Furthermore, the best results are obtained under assumption that the EFG has axial symmetry with the principal axis lying in the $[1\bar{2}1]$ plane. It appears that direction of the EFG principal axis coincides with either $\phi = 0$ or $\phi = \pi$, where one has $B(\phi = 0, \pi) = B_0 \exp[P_{20}] \approx B_0[1+P_{20}]$. Hence, the long axis of the cycloidal ellipse is aligned with the principal axis of the EFG as one finds $P_{20} > 0$. The quadrupole coupling constant $A_Q = \frac{1}{12}eQ_eV_{33}(c/E_0)$ was found positive as well indicating that the principal component of the EFG, i.e., $V_{33}$ is positive as the nuclear spectroscopic quadrupole moment $Q_e$ in the first excited state of the $^{57}Fe$ is positive ($+0.17$ b [11]). Here remaining symbols denote, respectively: $e$ the positive elementary charge, $c$ the speed of light in vacuum, and $E_0$ the energy of the resonant transition. A texture coefficient $g_{11}^{(111)}$ has been fitted to the major site [12], however it remained close to unity indicating non-significant sample orientation. Due to the smallness of the electric quadrupole interaction in comparison with the magnetic dipole interaction off-diagonal terms in the hyperfine Hamiltonian of the excited state are very small. This smallness prevents evaluation of other (mixing) texture coefficients for this dipolar nuclear transition. It is likely that the principal axis of the EFG is aligned with the $\langle 111 \rangle$ direction (electric polarization vector). Hence, the longer axis of the cycloidal ellipse is aligned with the same direction due to positive alignment of the residual orbital contribution with the spin induced Fermi field.

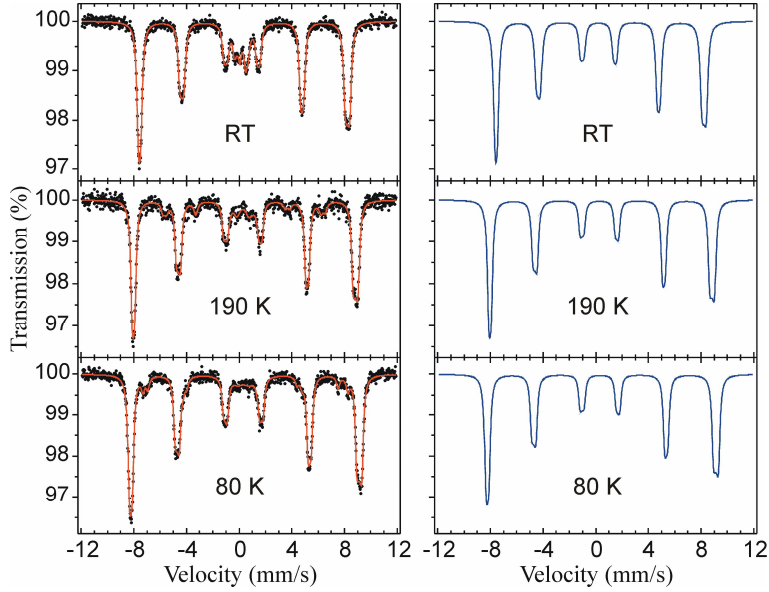

**Figure 1** Mössbauer transmission spectra obtained at three different temperatures are shown at left column with a continuous line being fitted pattern to the experimental data. Right column shows corresponding calculated spectra upon having removed all impurity phases.

**Table 1**

Essential hyperfine parameters versus temperature T. Symbols $C_1$, $C_2$, $C_3$, and $C_4$ stand for relative contributions to the total absorption cross-section with the index "1" marking major $BiFeO_3$ phase. Symbols $S_1$, $S_2$, $S_3$, and $S_4$ denote total shifts versus room temperature $\alpha-Fe$. Symbols $A_{Q1}$, $A_{Q2}$, $A_{Q3}$, and $A_{Q4}$ stand for the quadrupole coupling constants (see text for details). Symbols $B_3$ and $B_4$ stand for the hyperfine magnetic fields on impurities sites with magnetic order. Impurity "2" does not order magnetically within investigated temperature range. Symbols $\Gamma_1$ and $\Gamma_{234}$ stand for the absorber linewidths. The symbol $g_{11}^{(111)}$ denotes (dipolar – diagonal) texture parameter for the iron site of the major phase, i.e., for $BiFeO_3$. The symbol $B_0$ stands for the scaling field, while the symbol $P_{20}$ for the cycloid ellipticity. Errors for all values are of the order of unity for the last digit shown.

| BiFeO₃ (89 %) $C_1$ = 89 % | | | | | | |
|---|---|---|---|---|---|---|
| T (K) | $S_1$ (mm/s) | $A_{Q1}$ (mm/s) | $B_0$ (T) | $P_{20}$ | $\Gamma_1$ (mm/s) | $g_{11}^{(111)}$ |
| RT | 0.388 | +0.079 | 48.70 | +0.013 | 0.18 | 1.05 |
| 190 | 0.456 | +0.082 | 52.06 | +0.012 | 0.15 | 1.02 |
| 80 | 0.501 | +0.085 | 53.60 | +0.010 | 0.15 | 1.01 |

| Impurities (11 %) $C_2$ = 2 %, $C_3$ = 5 %, $C_4$ = 4 % $\Gamma_{234}$ = 0.2 – 0.3 (mm/s) | | | | | | | |
|---|---|---|---|---|---|---|---|
| T (K) | $S_2$ (mm/s) | $S_3$ (mm/s) | $S_4$ (mm/s) | $A_{Q2}$ (mm/s) | $A_{Q3}$ (mm/s) | $A_{Q4}$ (mm/s) | $B_3$ (T) | $B_4$ (T) |
| RT | 0.29 | 0.26 | 0.38 | 0.13 | 0.17 | 0.07 | 0 | 0 |
| 190 | 0.39 | 0.34 | 0.44 | 0.15 | +0.04 | +0.01 | 36.6 | 38.4 |
| 80 | 0.43 | 0.35 | 0.49 | 0.14 | +0.01 | +0.03 | 44.9 | 48.3 |

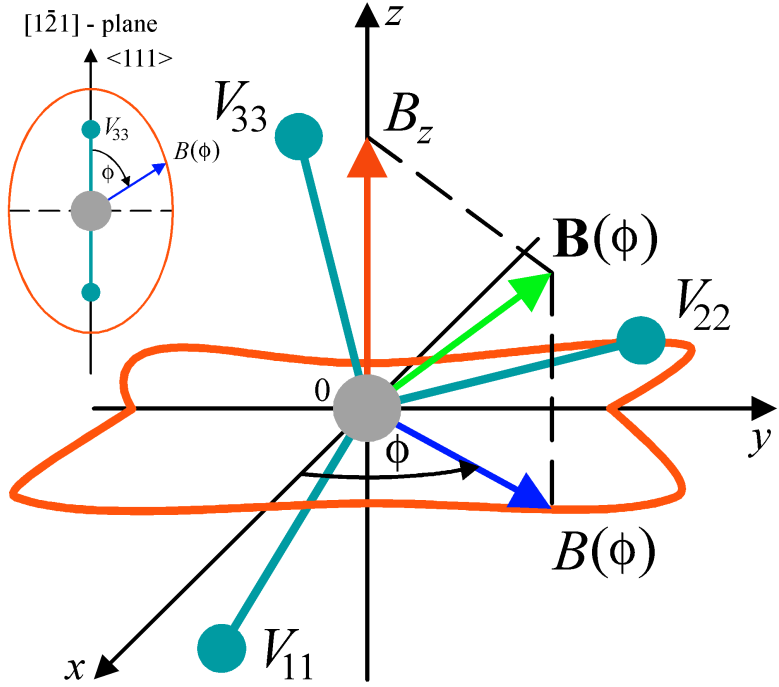

**Figure 2** General situation dealt with the application GmFeAs – see text for details. The hyperfine magnetic field "rotates evenly" in the [*xy*] plane being modulated along the trajectory circumference with the allowance for the constant perpendicular component. The EFG principal components are oriented arbitrarily in the reference frame {*xyz*} making another orthogonal system. Inset shows approximately the situation for $BiFeO_3$. The resonant nucleus stays in the origin.

General situation for either magnetic spiral or cycloid is shown in Figure 2. The nucleus is placed in the origin of the right-handed Cartesian system {*xyz*}. The spiral field $B(\phi)$ propagates in the [*xy*] plane. Some additional perpendicular component $B_z$ could appear for conical arrangements (here absent). Hence, the total field appears as **B**($\phi$) on the nucleus. The EFG tensor has three mutually orthogonal principal components $V_{11}$, $V_{22}$ and $V_{33}$, and it is traceless. The angle $\phi$ is the azimuthal angle of the Cartesian system. The principal components of the EFG are rotated from the Cartesian system by three Eulerian angles $\{\phi_1 \phi_2 \phi_3\}$ being respectively, the third Eulerian angle, polar angle, and azimuthal angle. For unrotated EFG one has $\{\phi_1 = 0 : \phi_2 = 0 : \phi_3 = 0\}$ with the subsequent principal components of the EFG $V_{11}$, $V_{22}$ and $V_{33}$ aligned with respective Cartesian axes {*xyz*}. The shape of the spiral is parameterized as [13]:

$$B(\phi) = B_0 \exp\left[\sum_{l=1}^{L}\sum_{m=0}^{l}\left(\frac{l!}{(l-m)!m!}\right)P_{lm}\cos^{l-m}(\phi)\sin^{m}(\phi)\right].$$

(1)

Usually, the index $L$ does not exceed four. For the case of $BiFeO_3$ (see, inset of Figure 2) one has planar cycloid with $B_z = 0$, axially symmetric EFG with $V_{11} = V_{22}$, Eulerian angles $\{\phi_1 = 0 : \phi_2 = \pi/2 : \phi_3 = 0\}$, scaling field $B_0 > 0$, and the sole non-trivial shape parameter $P_{20} > 0$, albeit very small leading to the linear approximation of the expression (1). It appears that one has $V_{33} > 0$, and rather small, as it is generated outside the iron atomic shell of electrons. Almost sinusoidal modulation of the hyperfine magnetic field along the ellipse circumference leads to the characteristic field distribution with sharp and diverging edges having smoother outlook from inside a distribution [14]. Similar distribution was reported in [7], and later rediscovered and reported in [6, 8]. This effect could be called anharmonicity of the cycloid. For example the hyperfine field at room temperature varies between 48.70 T and 49.34 T, while circumnavigating the cycloid. The total modulation depth being about 0.64 T at room temperature diminishes to about 0.54 T at 80 K. Apparent magnetic hyperfine field distribution could be calculated in the following way. The field belongs to the range $0 < B_0 \leq B(\phi) \leq B_1 = B_0 \exp(P_{20}) \approx B_0(1 + P_{20})$ with $P_{20} > 0$ in our case. One can calculate the angle $\pi/2 > \phi > 0$ as $\phi = a\cos\left\{\sqrt{P_{20}^{-1} \ln[B(\phi)/B_0]}\right\}$. Hence, a probability density function for the field $B$ distribution takes on the form $\rho(B) = N^{-1}(\partial \phi / \partial B)$ with $N = \int_{B_0}^{B_1} dB\,(\partial \phi / \partial B)$. A probability density function beyond the range $[B_0 : B_1]$ is equal zero, while it is positive within above range with the integral over the range being unity. One can even obtain a closed-form expression as $\rho(B) = \left\{\pi B \sqrt{\ln(B/B_0)[P_{20} - \ln(B/B_0)]}\right\}^{-1}$ or in the linear approximation, i.e. for very small parameter $0 < P_{20} \ll 1$ as $\rho(B) \approx B_0 \left\{\pi B \sqrt{(B - B_0)[B_0(1 + P_{20}) - B]}\right\}^{-1}$. The function $\rho(B)$ is shown in Figure 3 for temperatures of measurements. The function $\rho(B)$ diverges for $B = B_0$ and $B = B_1$, but this is weak divergence, and the function remains integrable. The poles at the ends of the physically meaningful argument are responsible for the fact, that the spectrum could be reasonably fitted with two iron "sites" having equal populations, differing by the magnetic hyperfine field and quadrupole coupling constant accounted for in the first order approximation. A difference in the apparent quadrupole coupling constant (including sign) is due to the fact, that for the lower field the orientation of the axial EFG principal axis is perpendicular to the magnetic field, while for the higher field this axis is aligned with the field. A small difference in the shape of the right and left cusp is primarily due to the non-linearity of the applied exponential scaling, and it is hardly visible in the experiment due to the smallness of the parameter $P_{20}$ and rather large value of the field $B_0$. The exponential scaling of equation (1) stems from the peculiarity of radial functions in the vicinity of the origin. A field distribution is *quasi*-continuous as the phase at the beginning of the cycloid is set rather randomly by the defect originating particular set of cycloids.

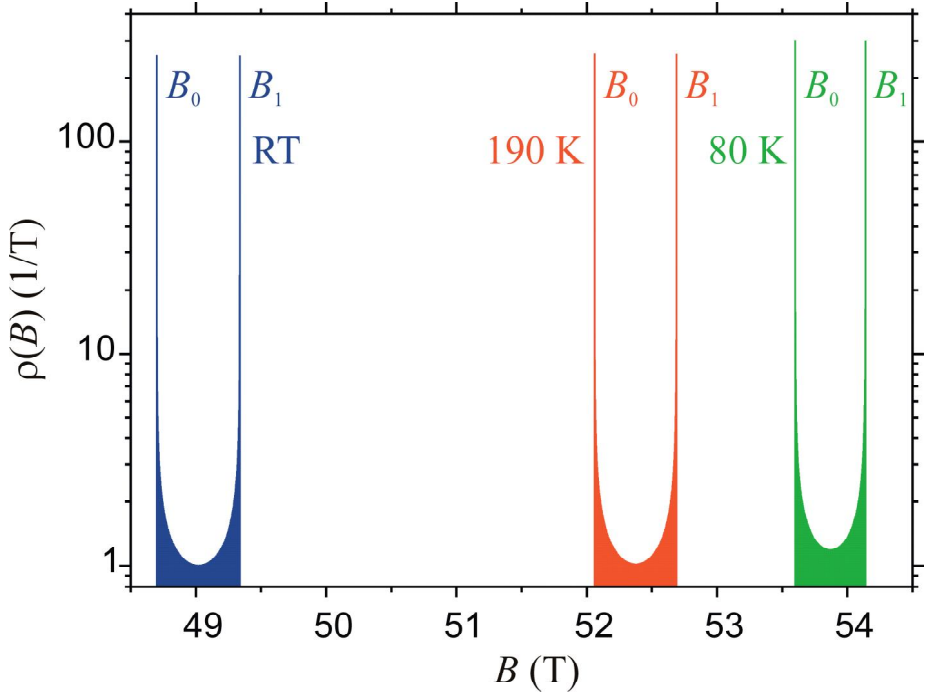

**Figure 3** Probability density of apparent field distribution plotted versus hyperfine field $B$ for three different temperatures. Fields $B_0$ and $B_1$ are marked for each temperature.

## 4. Conclusions

It was found that the EFG in BiFeO$_3$ is axially symmetric on the iron site with the principal axis aligned with the electric polarization vector, i.e. ⟨111⟩ direction of the rhombohedral chemical unit cell. The principal component of the EFG is positive. Attempt to fit asymmetry parameter of the EFG tensor yields definitely poorer results.

The magnetic cycloid is planar with the magnetic moments confined to the crystal plane [1$\bar{2}$1]. It is not perfectly circular, but it has small elliptic-like deformation. The longer axis of the ellipse is aligned with the crystal direction ⟨111⟩, while the shorter axis is very close to the cycloid propagation direction (it is exactly this direction provided rhombohedral distortion is neglected). Such deformation is in general agreement with previously found hyperfine field distributions [6-8]. A deformation of the magnetic cycloid diminishes with lowering temperature indicating that some excited electronic levels contribute to the cycloid ellipticity via the spin-orbit coupling. It was recently demonstrated that strong external magnetic field is able to deform the cycloid and eventually align iron magnetic moments [15]. Hence, the cycloid is fragile due to the weakness of the Dzyaloshinskii-Moriya interaction, and it could be deformed in the elliptic-like manner by even weak magnetostrictive forces following the spin-orbit coupling.